# Ultra-Fast 3D Porous Media Generation: a GPU-Accelerated List-Indexed Explicit Time-Stepping QSGS Algorithm

Ruofan Wang, Mohammed Al Kobaisi*

Department of Chemical and Petroleum Engineering, Khalifa University, Abu Dhabi, UAE
* Corresponding author: Mohammed.alkobaisi@ku.ac.ae

## Abstract

Efficient generation of high-resolution synthetic microstructures is essential in digital rock physics, yet classical Quartet Structure Generation Set (QSGS) algorithms become prohibitively expensive on large three-dimensional grids. We develop a list-indexed explicit time-stepping (LIETS) formulation of QSGS that restricts stochastic growth operations to an explicit active front instead of the entire voxel grid. The method is implemented in Python using NumPy on CPUs and CuPy on GPUs, and incorporates seed-spacing control via diamond dilation together with a volume-fraction-dependent directional growth probability. For a $400^3$ domain, LIETS reduces generation time from tens of minutes for a serial CPU implementation and several minutes for vectorized CPU and GPU QSGS to about 24 s on a consumer-grade RTX 4060, achieving peak throughputs up to $2.7\times10^7$ nodes/s. A Fontainebleau sandstone benchmark at $500^3$ resolution shows that LIETS reproduces the dependence of pore and grain size distributions on seed spacing (optimal $s$=30 voxels) and yields permeability-porosity trends within the experimental envelope and consistent with previously published Fast-QSGS results.

## 1. Introduction

Pore-scale structure fundamentally governs the transport, mechanical, and reactive properties of geomaterials and engineered porous media [1–5], and digital rock physics (DRP) has become a standard complement to laboratory core analysis for quantifying these relationships over the last few decades [6–8]. Recent DRP workflows couple high-resolution micro-CT imaging, segmentation, and direct numerical simulation with data-driven analysis, linking rock physics, imaging, and machine learning in unified frameworks [9–14]. Micro-CT-based digital rocks now routinely support the extraction of petrophysical and elastic properties [3,15,16], multiscale reconstruction across several orders of magnitude in length scale [11,17,18], and the construction of benchmark volumes for sandstones, carbonates, shale [19–25], and synthetic rocks [26–28], often augmented by laboratory porosity-permeability measurements.

Direct imaging at the resolution and field of view required for multi-scale flow and transport studies, however, remains costly and time-consuming, and it is often impractical to scan the large volumes and multiple realizations needed for uncertainty quantification. Synthetic microstructure generators therefore play a central role in DRP workflows, and a wide range of reconstruction strategies has been developed. Process-based approaches emulate sedimentation, compaction, and diagenesis to reconstruct sandstones directly from petrographic observations and limited statistical descriptors, often achieving close agreement with micro-CT images and measured transport properties [1,29]. Other frameworks adjust or optimize microstructures to match target petrophysical responses or geometric indicators, for example by guiding 3D reconstruction with bulk petrophysical data, enforcing strong pore connectivity, or explicitly accounting for large-scale anisotropy [30,31]. Recent stochastic reconstruction schemes based on simulated annealing and related energy-minimization methods remain widely used and have been systematically analyzed and extended to multiscale settings [32–35]. Multiple-point statistics (MPS) methods represent another important class, in which training images encode higher-order spatial correlations [36–39]; Single normal equation simulation (SNESIM)-type algorithms and their GPU-accelerated variants have been successfully applied to the reconstruction of 3D porous media and geological structures [40–42]. More recently, deep-learning-based approaches and supervised-learning schemes have been proposed to reconstruct, emulate, or design porous microstructures directly from images or statistical descriptors, effectively expanding the design space beyond what is easily accessible by manual parameter tuning [43–47].

Within this broad landscape, the quartet structure generation set (QSGS) stands out as a versatile stochastic generator. Originally introduced by Wang and co-workers as a four-parameter random growth algorithm to reproduce microscale random porous media and functionally graded materials, QSGS describes the microstructure in terms of a seed probability and three directional growth probabilities, and was successfully coupled with lattice Boltzmann solvers for predicting effective thermal and electrical conductivities [48,49]. Since then, QSGS has been applied to a wide range of porous systems, including random pore structures in particulate filters [50], soils with fractal micro-pore networks [51], rock porous media [52], and reconstructed porous concrete [53], enabling systematic studies of permeability and representative elementary volume (REV) behavior. Extensions of the method have been used to analyze pore-scale flow and solute transport heterogeneity induced by mineral dissolution and precipitation in multiphase rocks [54,55]. QSGS-generated random isotropic porous media have also served as a basis for evaluating effective thermal conductivity over controlled ranges of porosity and pore size, while parametric studies have quantified how the four growth parameters influence pore morphology and connectivity in generic porous materials [56].

Despite its conceptual simplicity and flexibility, the computational cost of classical

QSGS implementations remains substantial for large three-dimensional voxels. In typical serial codes, each growth iteration sweeps the entire domain, generating random numbers and evaluating neighborhood rules at every grid cell, even though only sites in the vicinity of the growth front can actually change phase. This full-field update pattern leads to per-step work that scales with the total number of cells, independent of the current solid volume fraction, and results in tens of minutes of wall-clock time when generating microstructures on hundreds-cubed voxels. Recognizing this bottleneck, recent work has exploited vectorization and GPU acceleration: the Fast-QSGS program recasts the growth stage as array operations in a NumPy-compatible style and executes them on modern GPUs via CuPy [57–59], achieving order-of-magnitude speedups over serial implementations on both consumer-grade and data-center GPUs. Nevertheless, these vectorized full-field schemes still apply random number generation, logical masking, and voxel shifts to every cell for every direction at each global iteration, so a large fraction of compute and memory traffic is expended on sites that are far from the evolving interface and cannot grow. This limitation is particularly pronounced on consumer GPUs with modest memory bandwidth, where global memory access dominates runtime. There is therefore a clear need for algorithms that preserve the stochastic growth statistics and parameterization of QSGS while restricting expensive operations to the active growth front and remaining compatible with high-level array programming frameworks used in DRP workflows.

In this work, we develop a list-indexed explicit time-stepping (LIETS) formulation of the QSGS algorithm that concentrates all heavy computation on the active growth front. In the proposed scheme, the current interface is represented by an explicit list of active sites; for each global iteration and each voxel direction, only neighbors of these sites are considered as growth candidates, and newly occupied cells are appended to the active list for the next time step after pruning out fully surrounded sites. This list-based growth rule preserves the stochastic characteristics of QSGS, including seed spacing control and volume-fraction-dependent directional growth probabilities, while dramatically reducing the number of random draws and array operations. The LIETS formulation is implemented in Python using NumPy on CPUs and CuPy on GPUs, providing a compact, high-level code base that is portable across consumer-grade and professional hardware. Systematic performance benchmarks demonstrate that, for a $400^3$ voxel, the LIETS implementation reduces generation time from tens of minutes for serial QSGS and several minutes for vectorized CPU and GPU QSGS to tens of seconds on a consumer RTX 4060, and achieves peak throughputs on the order of $10^7$ nodes/s for moderately sized domains. Using Fontainebleau sandstone as a benchmark medium, we further show that LIETS-generated microstructures reproduce the expected dependence of pore and grain size distributions on seed spacing, and that permeability-porosity data obtained from pnflow simulations on these structures fall within the experimental envelope and agree well with previously published Fast-QSGS results. Together, these results demonstrate that the LIETS QSGS algorithm delivers substantial algorithmic performance gains while preserving physical realism, and that high-resolution microstructure generation for granular disordered media can be performed efficiently on mainstream desktop hardware.

## 2. Model and method

### 2.1 Basic QSGS procedure

In the basic Quartet Structure Generation Set (QSGS) method [48], the microstructure is discretized on an orthogonal grid, and all phases are generated in sequence as follow:

1) *Phase assignment.* One phase is selected as the matrix (non-growing) phase. All other phases are treated as growing phases and are generated one after another; the currently treated phase is referred to as the $n$-th phase.
2) *Seeding of the first growing phase.* A uniform random number in (0,1) is assigned to every grid cell. Cells whose random number is not larger than the prescribed core (seed) probability $c_d$ of the first phase are marked as cores of that phase. The value of $c_d$ is chosen not greater than the target volume fraction of the phase.
3) *Growth of the first phase.* Starting from all cells belonging to the first phase, growth is attempted towards all neighboring cells. For a neighbor located in direction $i$, a new random number is drawn; the neighbor is converted to the growing phase if this number does not exceed the directional growth probability $D_i$. Newly occupied cells are added to the set of growth sites for the next iteration. These growth iterations are repeated until the volume fraction of the first phase reaches its prescribed value (for a pore or gas phase this is often expressed in terms of the porosity). A concrete serial implementation of this growth procedure for a single phase in a two-dimensional grid is summarized in **Algorithm 1**.
4) *Subsequent growing phases.* The remaining phases are generated on the still unassigned cells. If the new phase behaves as an independent discrete phase, its cores are placed and grown in the same way as for the first phase, using its own core probability and growth probabilities. If the new phase interacts with existing phases, growth into a neighbor that currently belongs to phase $m$ occurs with a phase-interaction probability $I_i^{n,m}$, which specifies the chance that the $n$-th phase grows along direction $i$ at an interface with phase $m$. Growth of the $n$-th phase is stopped once its prescribed volume fraction $P^n$ is attained.
5) *Assignment of the matrix phase.* After all growing phases have been generated, any grid cell that has not been occupied is labelled as the matrix (non-growing) phase. The resulting phase map constitutes the synthetic microstructure.

**Algorithm 1**
Serial Growth Process
---
1: **while** $\phi_a < \phi_s$
2:    **for each** $x, y$ **in** $domain(Nx, Ny)$
3:       **if** $phase(x, y) = 1$
4:          **for each** $d$ **in** $directions$
5:             $nx = x + ex_d$ , $ny = y + ey_d$
6:             **if** $(nx, ny)$ **in** domain **and** $phase(nx, ny) = 0$ **and** $rand\_num < D_d$
7:                $phase(nx, ny) = 1$

```
8:         end if
9:       end for
10:     end if
11:  end for
12:  $\phi_a$ = volume_fraction(phase = 1)
13: end while
```

## 2.2 Vectorized QSGS procedure

To accelerate QSGS, the growth stage can be reformulated as array operations, which are well suited for vector processors and GPUs [57]. For clarity, consider a single solid phase in a two-dimensional grid; solid cells are encoded by 1 and void cells by 0. The corresponding vectorized growth procedure is outlined below and its implementation is summarized in **Algorithm 2**.

1) *Initial state*. Seeds of the solid phase are already placed according to a given seed probability $S_d$. Let $\phi_a$ be the current solid volume fraction and $\phi_s$ be the target value.
2) *Mask construction*. While $\phi_a < \phi_s$, a growth iteration is performed. A Boolean array *solid_mask* is created that is 1 at all current solid cells and 0 elsewhere.
3) *Direction-wise candidate generation*. For each voxel direction $i$ (e.g. up, down, left, right, and possibly the diagonals):
   - Draw a random array $R_i$ with entries uniformly distributed in (0,1).
   - Form a candidate array by testing $R_i \leq G_i$, where $G_i$ is the growth probability in direction $i$.
   - Restrict potential growth to the vicinity of existing solid cells by taking the logical AND of this candidate array with *solid_mask*.
   - Shift (roll) the resulting array by one grid cell in direction $i$; the shifted array indicates which neighbor cells will be filled from that direction.
4) *Update of the solid phase*. The shifted arrays from all directions are combined using the logical OR operation to obtain the set of cells to be converted to solid in this iteration. The solid array is updated accordingly, and the new volume fraction $\phi_a$ is computed. Cells that receive growth attempts from several directions are simply counted once.
5) *Iteration*. Steps 2-4 are repeated until $\phi_a$ reaches $\phi_s$. In this vectorized scheme, newly formed solid cells participate as seeds only in the next global iteration, rather than immediately after their creation as in a strictly serial implementation; because growth is strongly stochastic, the resulting structures remain statistically equivalent. A concise pseudocode implementation of the vectorized procedure for a single phase is provided in **Algorithm 2**.

**Algorithm 2**
Vectorized Growth Process

```
 1: while  $\phi_a < \phi_s$
 2:    solid_mask = (phase = 1)
 3:    for each  d  in directions
 4:       growth = **rand**(Nx, Ny) < $G_d$
 5:       growth = growth * solid_mask
 6:       phase = phase OR **shift2D**(growth, $ex_d$, $ey_d$)
 7:    end for
 8:    phase = (phase ≠ 0)
 9:    $\phi_a$ = volume_fraction(phase = 1)
10: end while
```

**2.3 List-indexed vectorized explicit time stepping QSGS**

To further reduce redundant operations in the growth stage, the QSGS method can be reformulated in terms of an explicit list of active growth sites instead of operating on the full grid. As in the vectorized formulation, consider a single solid phase on a two-dimensional grid, with solid cells encoded by 1 and void cells by 0. The main steps of the list indexed growth procedure are described below and concisely presented in **Algorithm 3**.

1) *Initial state*. Seeds of the solid phase are already placed according to a given seed probability $S_d$. Let $\phi_a$ denote the current solid volume fraction and $\phi_s$ the target value. The solid-void map is stored in an array $(x, y) \in \{0,1\}$.

2) *Active front construction*. At the beginning of the list-indexed procedure, an active set $A$ is constructed that contains all solid cells which have at least one neighboring void cell. Each element of $A$ is therefore a potential source for further growth. The voxel directions are the same as in the basic and vectorized QSGS formulations, and the directional growth probabilities are denoted by $G_i$ for direction $i$.

3) *List-based directional growth*. While $\phi_a < \phi_s$ and the active set $A$ is not empty, one growth iteration is performed. For each voxel direction $i$, all neighbors of the current active sites in that direction are collected, restricted to cells that lie inside the domain. From this set, only those neighbors that are still void are kept as candidates. For each candidate cell in direction $i$, an independent random number in $(0,1)$ is drawn; the candidate is converted to solid if this number does not exceed the growth probability $G_i$. The accepted cells from direction $i$ are added to a temporary set $A_{\text{new}}$ and their entries in the phase array are set to 1.

4) *Update of the active front*. After all directions have been processed, the sets of newly occupied cells from all directions are combined to form $A_{\text{new}}$. This set is then pruned: duplicate entries are removed and cells that no longer have any void neighbors are discarded, so that only sites that can still support further growth remain active. The resulting set replaces the previous active set, $A \leftarrow A_{\text{new}}$, and the new solid volume fraction $\phi_a$ is computed from the updated phase array.

5) *Iteration*. The list-indexed growth iterations are repeated until $\phi_a$ reaches $\phi_s$ or the active set becomes empty. In this scheme, newly formed solid cells enter the active set only in the next global iteration, similar in spirit to the vectorized procedure, but the operations are restricted to the vicinity of the growth front rather than to the full grid. The corresponding list based algorithm for a single phase is summarized in **Algorithm 3**.

A list indexed vectorized explicit time stepping QSGS formulation has a clear computational advantage over a full field vectorized scheme because it restricts all heavy operations to the active growth front instead of the entire voxel. In the standard vectorized QSGS procedure, each growth iteration applies random number generation, logical masking and voxel shifts to all grid cells in $\Omega$ for every voxel direction $i$, even though only a small subset of cells near existing solid regions can actually change state. The per step cost therefore scales with the total number of cells $N_{\text{cells}}$ multiplied by the number of directions, irrespective of the current solid volume fraction $\phi_a$. In contrast, the list indexed formulation maintains an explicit active set $A$ that contains only solid cells with at least one void neighbor. Direction wise neighbor construction, probability tests with $G_i$, and phase updates are performed only for this set and its neighbors. The per step cost scales with the size of the growth front, $N_{\text{active}}$, which is typically much smaller than $N_{\text{cells}}$s except in the very early stages of growth.

This reduction in the number of sites processed has several secondary benefits at the implementation level. First, random numbers are generated only for candidate neighbors in $C_i$, not for the whole domain, which lowers both arithmetic and memory traffic. Second, memory access patterns are more localized: reading and writing phase values is concentrated around the evolving interface, which improves cache reuse on CPUs and reduces wasted bandwidth on GPUs. Third, the explicit time stepping structure ensures that newly created solid cells are appended to the active set only once per global iteration, avoiding repeated work on cells that can no longer grow. As a result, for typical target fractions $\phi_s$ and realistic directional probabilities $G_i$, the list indexed scheme achieves the same stochastic growth statistics as the full field vectorized QSGS, but with a substantially lower number of array operations and random draws per iteration, which translates into shorter runtimes especially on large and moderately sparse grids.

In order to ensure a fair comparison and to isolate algorithmic effects from parameter choices, the parameter settings in this work are chosen to be consistent with those reported by Yang et al [57]. They introduced a seed-spacing parameter $s$ that defines the minimum allowable distance between seeds, thereby improving the spatial uniformity of the initial nuclei distribution. Several strategies for enforcing seed spacing were examined in their work. This study adopts the same spacing strategy selected by Yang et al., namely the regular diamond dilation scheme [57]. In this approach, the dilation shape corresponds to a square in two dimensions and an octahedron in three dimensions, rather than a sphere, as illustrated in **Fig. 1**. This choice is motivated by computational efficiency: diamond-shaped regions can be generated through successive dilation operations using a simple-connectivity kernel, which is significantly less expensive than constructing spherical neighborhoods. In addition, the present work follows Yang et al. [57] by employing a variable directional growth probability $G_i(\phi_a)$ that depends on the current volume fraction, instead of the constant growth probability used in the original QSGS formulation. The corresponding functional form is given by

$$G_i(\phi) = G_{i,ref} \frac{\phi_s - 0.95\phi_a}{0.05\phi_s} \qquad (1)$$

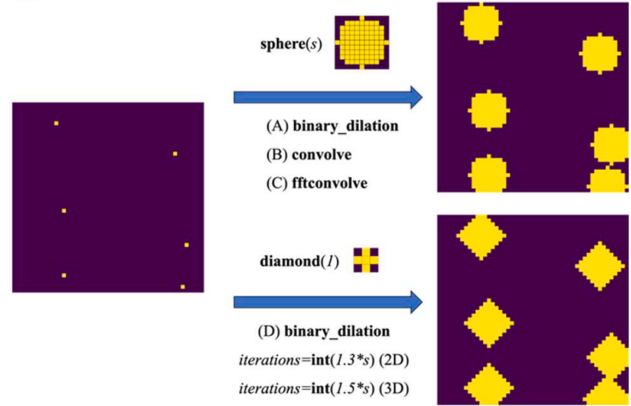

**Fig. 1**. Illustration of different seed spacing methods: (A) sphere dilation; (B) direct convolve; (C) FFTconvolve; (D) diamond dilation [Yang et al. [57]]. For the first three methods, the dilation is conducted with a large kernel, while diamond dilation is conducted with a small kernel but multiple iterations. The volume (area) of the diamond is equal to that of the sphere.

**Algorithm 3**
List-indexed Vectorized Growth Process

1:   $A = \{(x, y) \text{ in } domain(Nx, Ny) \mid phase(x, y) = 1\}$
2:   **while** $\phi_a < \phi_s$ **and** $A$ **not** *empty*
3:     $A_{new} = empty\ set$
4:     **for each** $d$ in *directions*
5:       $N_d = neighbors(A,\ ex_d,\ ey_d)$ **inside** *domain*
6:       $C_d = \{q \text{ in } N_d \mid phase(q) = 0\}$
7:       **for each** $q$ in $C_d$
8:         **if** $rand\_num < G_d$
9:           $phase(q) = 1$
10:           $A_{new} = A_{new} \cup \{q\}$
11:         **end if**
12:       **end for**
13:     **end for**
14:     $A = prune(A_{new},\ phase)$
15:     $\phi_a = volume\_fraction(phase = 1)$
16:   **end while**

## 3. Results and discussion

### 3.1. Computational platform and hardware configuration

All computations in this study were carried out on a standard consumer grade desktop rather than on professional workstations, due to practical constraints. The system consists of an Intel Core i7 14700KF CPU, 64 GB of DDR5 memory, and an NVIDIA GeForce RTX 4060 GPU. The RTX 4060 is based on the Ada Lovelace architecture and provides a peak FP32 performance of 15.11 TFLOPs and a peak FP64 performance of 236.2 GFLOPs, with 8 GB of GDDR6 memory and a peak memory bandwidth of 272 GB/s; it is a mainstream model released in 2023 and does not match the memory capacity or bandwidth of typical high performance computing nodes. Under these hardware conditions, the performance improvements reported here arise from the efficiency of the algorithms described in Section 2, particularly the list indexed vectorized explicit time stepping QSGS formulation, rather than from any advantage in GPU specifications. In practice, the optimized implementation on this consumer grade platform attains generation times for high resolution porous structures that are comparable to or faster than those reported in studies using professional grade workstations or clusters, indicating that the observed speedup is primarily due to algorithmic design rather than access to high end computing resources.

### 3.2. Performance comparison and scalability analysis

To facilitate a direct and rigorous comparative analysis, the computational setup adopted in this study mirrors the benchmark configuration established by Yang et al. [57]. The simulation domain comprises a cubic voxel grid with a discretization resolution of $400^3$. The stochastic parameters governing the QSGS are defined as a seed probability of $S_d = 2 \times 10^{-4}$ and a reference directional growth probability of $G_{i,ref} = 8 \times 10^{-4}$, with the porosity $\phi$ fixed at 0.2. For all validation cases, a seed spacing parameter of *s*=15 is imposed. A diamond-type dilation scheme is employed to enforce this constraint, utilizing repeated dilation with a diamond-shaped kernel rather than spherical neighborhoods to optimize memory access patterns.

The computational efficiency of the evaluated implementations, utilizing single-precision floating-point arithmetic (FP32), is elucidated in **Fig. 2(a)**. The baseline serial algorithm, executed on the CPU, exhibits significant computational overhead, requiring approximately 23.5 minutes to generate a single realization on the $400^3$ voxel. Leveraging vector operations within a CPU parallel framework reduces this runtime to roughly 5 minutes. When ported to the GPU, the vectorized QSGS implementation maintains a runtime of approximately 6 minutes. In stark contrast, our LIETS algorithm completes the structure generation in merely 24 seconds, providing speedup factors of approximately 60 compared to the CPU serial code, 12 compared to the CPU vectorized version, and 15 compared to the GPU vectorized implementation. To contextualize these findings, **Fig. 2(b)** juxtaposes our results with the runtime metrics reported in [57]. While the conventional QSGS-based algorithm necessitates around 100 seconds on an RTX 4060 GPU and approximately 30 seconds on a data-center class A100 GPU for comparable $400^3$ cases, our LIETS algorithm demonstrates superior efficiency. Remarkably, despite being deployed on a consumer-grade RTX 4060, our LIETS algorithm achieves a structure generation speed that is comparable to, and in some instances exceeds, the speed achieved on the high-end A100 hardware. Collectively, these benchmarks corroborate that the LIETS algorithm delivers a substantial leap in structure generation efficiency over existing QSGS implementations.

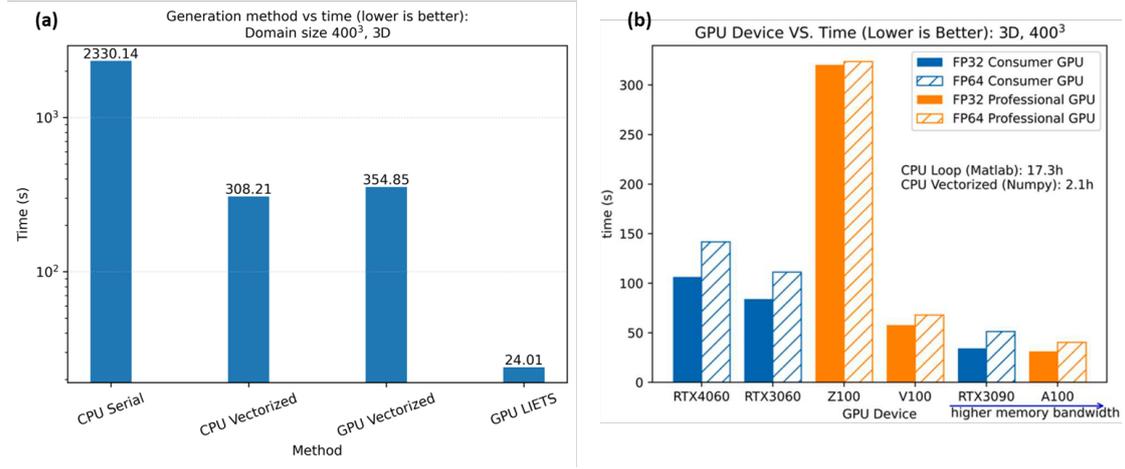

**Fig. 2**. Overall runtime for generating a 3D QSGS microstructure with domain size $400^3$. (a) Wall clock time of the CPU serial, CPU vectorized, GPU vectorized, and GPU LIETS implementations on an RTX 4060 using FP32 arithmetic. (b) GPU runtime of the QSGS implementation of Yang et al. [57] on different consumer and professional devices, shown for comparison.

**Fig. 3** characterizes the computational throughput of the LIETS-based structure generator on an RTX 4060 GPU in 3D, quantified in nodes per second for both single precision (FP32) and double precision (FP64). **Fig. 3(a)** illustrates the performance under a seed spacing constraint ($s$=15) enforced via diamond-type dilation, whereas **Fig. 3(b)** depicts the unconstrained scenario. Under the spacing constraint $s$=15, the FP32 implementation achieves a speed ranging from $9.7 \times 10^6$ to $1.5 \times 10^7$ nodes/s for domain sizes between $200^3$ and $600^3$, peaking at $1.51 \times 10^7$ nodes/s for the $600^3$ domain. As the domain size expands further ($700^3$ to $1000^3$), a decline in computational speed is observed – dropping to between $6.1 \times 10^6$ and $2.1 \times 10^6$ nodes/s. This degradation is likely caused by memory bandwidth saturation and reduced cache effectiveness at larger voxel sizes. It may also reflect conservative power and thermal limits on the consumer grade RTX 4060 for a personal desktop setting. A similar trend is evident in the FP64 results, where the peak speed reaches $8.12 \times 10^6$ nodes/s at $500^3$ and remains robust (above $7.0 \times 10^6$ nodes/s) up to $600^3$. For the largest domains investigated, the FP64 speed converges to approximately $1.6 \times 10^6$ to $2.2 \times 10^6$ nodes/s. Quantitatively, the algorithmic overhead associated with enforcing seed spacing reduces the generation rate by approximately 50% compared to the unconstrained case; nevertheless, the method sustains a high speed on the order of $10^7$ nodes/s for moderate domain sizes.

Relieving the spacing constraint significantly enhances computational performance, as delineated in **Fig. 3(b)**. In the FP32 regime, the speed exhibits a monotonic enhancement from $1.41\times10^7$ nodes/s at $200^3$ to a maximum of $2.70\times10^7$ nodes/s at $800^3$. Even at the maximal resolution of $1000^3$, the FP32 speed is sustained at $8.01\times10^6$ nodes/s. Similarly, the FP64 implementation achieves a peak speed of $1.29\times10^7$ nodes/s at $500^3$, maintaining speeds above $7.0\times10^6$ nodes/s for domains up to $600^3$. For larger grids ($800^3$ to $1000^3$), the FP64 speed stabilizes in the range of $4.52\times10^6$ to $3.57\times10^6$ nodes/s. A comparative analysis of **Figs. 3(a)** and **3(b)** indicates that eliminating the spacing constraint amplifies the generation speed by a factor of 1.4 to 1.8 for domain sizes up to $500^3$, and by roughly 2 to 4 for larger domains where the computational intensity of neighbor-checking operations becomes more pronounced.

The external benchmark from Yang et al. [57] is recast in terms of generation speed and shown in **Fig. 4**. For the diamond-type QSGS generator with spacing $s=15$, the reference plateau speeds on an RTX 3060 are approximately $7.5\times10^5$ nodes/s (FP32) and $5.5\times10^5$ nodes/s (FP64), scaling to roughly $2.1\times10^6$ and $1.6\times10^6$ nodes/s on an A100 GPU. In comparison, our LIETS implementation on a consumer RTX 4060 achieves FP32 speeds of approximately $1.4\times10^7$ nodes/s and FP64 speeds up to $8.1\times10^6$ nodes/s for $400^3$ to $500^3$ domains. This represents a performance advantage of roughly one order of magnitude over the RTX 3060 results and a three- to seven-fold acceleration compared to the A100 data reported in [57]. Furthermore, without the spacing constraint, the LIETS generator attains peak speeds of $2.7\times10^7$ nodes/s (FP32) and $1.3\times10^7$ nodes/s (FP64). Overall, the comparative evidence from **Figs. 2** to **4** demonstrates that our algorithm consistently outperforms existing QSGS based generators on both consumer class and data center class GPUs, and thus provides a highly competitive computational tool for large scale microstructure generation in multi-phase porous media.

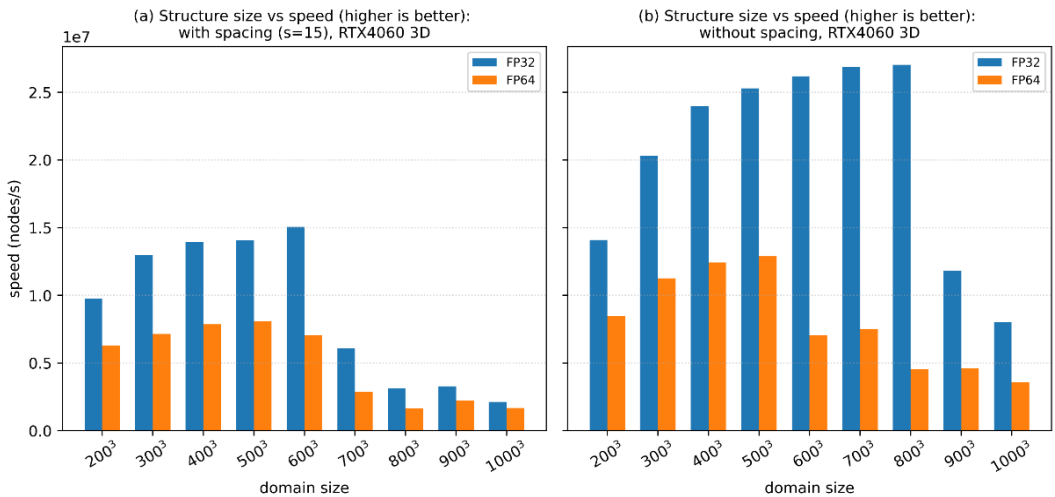

**Fig. 3**. Throughput of the LIETS-based structure generator on an RTX 4060 GPU as a function of 3D domain size, measured in nodes per second. (a) Results with a seed spacing constraint s=15 enforced by diamond-type dilation for both FP32 and FP64. (b) Corresponding results without seed spacing.

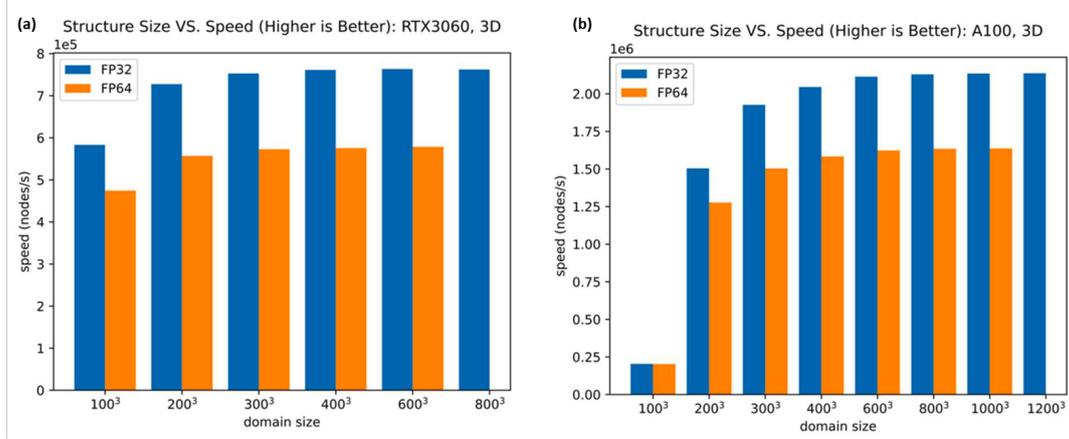

**Fig. 4**. Reference throughput of the diamond-type QSGS generator from Yang et al. [57] for varying 3D domain sizes, measured in nodes per second. (a) Performance on an RTX 3060 GPU in FP32 and FP64. (b) Performance on an NVIDIA A100 GPU in FP32 and FP64.

### 3.3. Benchmark test

Following Yang et al. [57], Fontainebleau sandstone is likewise selected as the benchmark medium for QSGS based structure generation. Fontainebleau sandstone is a well sorted, quartz dominated granular porous medium with negligible clay content, an intergranular porosity typically reported in the range 0.03~0.30, and an average grain diameter of 200~250 μm. In this benchmark, the solid seeds for the sandstone are prescribed using the same expression as Yang et al., $S_d = 6\phi_s \Delta^3 / \pi d^3$, where $\phi_s$ is the solid fraction, $\Delta = 4$ μm/voxel is the voxel size, and $d$ denotes the target grain diameter. The computational domain is a cube of size $500^3$ voxels, the porosity is fixed at 0.15, and the initial growth probability is set to $G_{i,ref} = 0.0008$. The seed spacing parameter $s$ is varied from 0 to 40 (i.e., 0 to 160 μm) with an increment of 5 voxels to examine its influence on the generated Fontainebleau type microstructures.

To obtain statistically representative results, five independent realizations are generated for each value of the spacing parameter $s$. The resulting digital rock samples are analyzed in PerGeos to extract the grain size and pore size distributions, and the outcomes are summarized in **Fig. 5**. The overall trends are fully consistent with the results of Yang et al. [57], although the quantitative values do not match exactly due to the use of different analysis tools. In particular, for $s$=30 the data exhibit the best convergence: the maximum pore size is the smallest among the nine spacing cases, and the grain size statistics show the clearest trend, with the smallest mean value and the narrowest range across all nine groups.

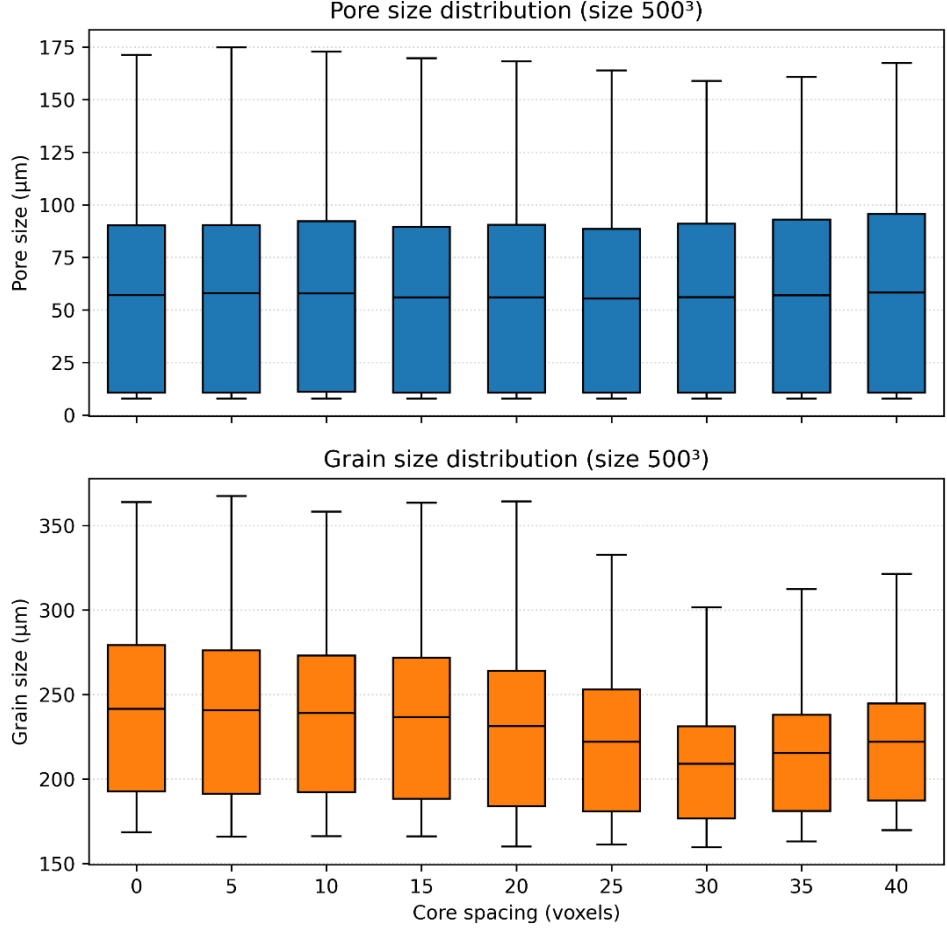

**Fig. 5**. Effect of seed spacing on pore and grain size distributions for QSGS generated Fontainebleau sandstone (domain size $500^3$); box plots of the pore size (top) and the grain size (bottom).

To validate the flow properties, fifteen digital rock samples with porosities ranging from 0.0542 to 0.2657 were generated using the LIETS-QSGS algorithm. Apart from the target porosity, all microstructural generation parameters were maintained consistent with the preceding benchmark. The intrinsic permeability of each realization was subsequently determined using the open-source solver pnflow, which performs flow simulation on pore networks extracted from the voxelized structures [60,61]. The computed permeability-porosity data were superimposed onto the experimental dataset for Fontainebleau sandstone compiled by Yang et al. as shown in **Fig. 6** [57,62–67], alongside their numerical predictions. The results from the present study fall entirely within the experimental envelope, demonstrating that the microstructures generated by LIETS exhibit transport properties consistent with laboratory observations. We note that the simulated permeabilities at $\phi = 0.10$ and 0.15 are marginally lower than the data reported by Yang et al. This minor discrepancy is likely attributed to the distinct numerical schemes employed: while Yang et al. utilized a direct voxel-scale Lattice Boltzmann Method (LBM), our study relies on pore network modeling (PNM). The geometric simplification inherent in network extraction and upscaling procedures is known to yield slightly more conservative permeability estimates.

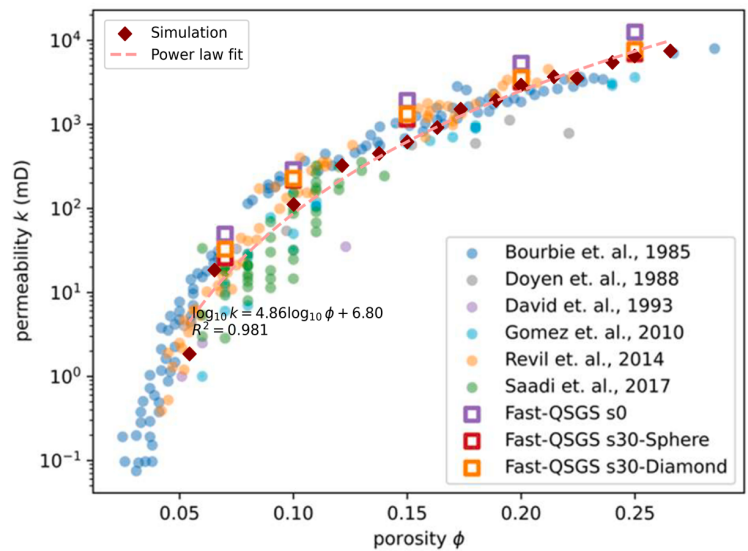

**Fig. 6**. Permeability-porosity relation for Fontainebleau sandstone. The colored circles and Fast-QSGS markers correspond to experimental data and QSGS based simulations compiled by Yang et al. [57,62–67], while the burgundy diamond symbols and dashed power law curve represent the permeability results obtained in this work, superimposed on the same dataset for direct comparison.

## 4. Conclusion

In this work, we developed a list indexed explicit time stepping (LIETS) vectorized formulation of QSGS and implemented it in Python using NumPy and CuPy. For a $400^3$ voxel, our LIETS algorithm reduces the generation time from tens of minutes for serial codes, and several minutes for a vectorized CPU and a conventional vectorized GPU QSGS, to only 24 s on a consumer grade RTX 4060, with peak throughputs up to $2.7 \times 10^7$ nodes/s for larger domains. A systematic benchmark on Fontainebleau sandstone shows that the method can reproduce the expected dependence of pore and grain size on seed spacing and yields an optimal spacing around $s = 30$ voxels. Permeability–porosity data obtained from simulations on LIETS generated samples fall entirely within the experimental envelope for Fontainebleau sandstone and agree well with previously published Fast-QSGS results [57]. These findings demonstrate that the LIETS QSGS algorithm delivers substantial performance gains while preserving physical realism, and that high resolution microstructure generation for granular disordered media can be performed efficiently even on modest consumer hardware.

**Data Availability Statement**: The code developed in this work is open-sourced and

available at https://github.com/RonKU42/LIETS-QSGS

**Acknowledgments**: Ruofan Wang acknowledges Khalifa University for providing scholarship for his graduate studies. The authors acknowledge the use of AI tools to assist with language refinement.

**Author contributions**: R. W.: Writing – original draft, Software, Investigation, Data curation, Conceptualization. M. A. K.: Writing – review & editing, Supervision, Investigation, Formal analysis, Conceptualization.

**Conflicts of Interest**: The authors declare no conflicts of interest.

# Reference

1. ØREN, P.-E. & Bakke, S. Process Based Reconstruction of Sandstones and Prediction of Transport Properties. *Transport in Porous Media* **46**, 311–343 (2002).
2. Zhao, Z. & Zhou, X.-P. Pore-scale diffusivity and permeability evaluations in porous geomaterials using multi-types pore-structure analysis and X-μCT imaging. *Journal of Hydrology* **615**, 128704 (2022).
3. Liang, J. *et al.* Elastic Moduli of Arenites From Microtomographic Images: A Practical Digital Rock Physics Workflow. *Journal of Geophysical Research: Solid Earth* **125**, e2020JB020422 (2020).
4. Wang, Z., Hu, M. & Steefel, C. Pore-Scale Modeling of Reactive Transport with Coupled Mineral Dissolution and Precipitation. *Water Resources Research* **60**, e2023WR036122 (2024).
5. Du, X., Liu, E., Shan, R., Yang, F. & Ling, B. Pore-scale flow and solute transport heterogeneity induced by mineral dissolution. *Physics of Fluids* **37**, 066602 (2025).
6. Mehmani, A., Kelly, S. & Torres-Verdín, C. Leveraging digital rock physics workflows in unconventional petrophysics: A review of opportunities, challenges, and benchmarking. *Journal of Petroleum Science and Engineering* **190**, 107083 (2020).
7. Huang, C. *et al.* Construction of pore structure and lithology of digital rock physics based on laboratory experiments. *J Petrol Explor Prod Technol* **11**, 2113–2125 (2021).
8. Verri, I. *et al.* Development of a Digital Rock Physics workflow for the analysis of sandstones and tight rocks. *Journal of Petroleum Science and Engineering* **156**, 790–800 (2017).
9. Tembely, M., AlSumaiti, A. M. & Alameri, W. S. Machine and deep learning for estimating the permeability of complex carbonate rock from X-ray micro-computed tomography. *Energy Reports* **7**, 1460–1472 (2021).
10. Wang, Z. (王梓) *et al.* Pore-scale study of mineral dissolution in heterogeneous structures and deep learning prediction of permeability. *Physics of Fluids* **34**, 116609 (2022).
11. Jiang, F. *et al.* Upscaling Permeability Using Multiscale X-Ray-CT Images With Digital Rock Modeling and Deep Learning Techniques. *Water Resources Research* **59**, e2022WR033267 (2023).
12. Xie, C. (谢驰宇) *et al.* Relative permeability curve prediction from digital rocks with variable sizes using deep learning. *Physics of Fluids* **35**, 096605 (2023).
13. Gärttner, S., Alpak, F. O., Meier, A., Ray, N. & Frank, F. Estimating permeability of 3D micro-CT images by physics-informed CNNs based on DNS. *Comput Geosci* **27**, 245–262 (2023).
14. Wang, Z., Hou, Z. & Cao, D. Deep-learning-based digital rock physics analysis: from image segmentation and edge detection by few-shot learning to mechanical properties prediction. *Geoenergy Science and Engineering* **256**, 214133 (2026).

15. Saxena, N. *et al.* Rock properties from micro-CT images: Digital rock transforms for resolution, pore volume, and field of view. *Advances in Water Resources* **134**, 103419 (2019).
16. Ahmad, R., Liu, M., Ortiz, M., Mukerji, T. & Cai, W. Computation of effective elastic moduli of rocks using hierarchical homogenization. *Journal of the Mechanics and Physics of Solids* **174**, 105268 (2023).
17. Tian, Y. *et al.* Digital rock modeling of deformed multi-scale media in deep hydrocarbon reservoirs based on in-situ stress-loading CT imaging and U-Net deep learning. *Marine and Petroleum Geology* **171**, 107177 (2025).
18. Lin, W. *et al.* Multiscale Digital Porous Rock Reconstruction Using Template Matching. *Water Resources Research* **55**, 6911–6922 (2019).
19. Mayo, S. *et al.* Quantitative micro-porosity characterization using synchrotron micro-CT and xenon K-edge subtraction in sandstones, carbonates, shales and coal. *Fuel* **154**, 167–173 (2015).
20. Pereira Nunes, J. P., Blunt, M. J. & Bijeljic, B. Pore-scale simulation of carbonate dissolution in micro-CT images. *Journal of Geophysical Research: Solid Earth* **121**, 558–576 (2016).
21. Callow, B., Falcon-Suarez, I., Marin-Moreno, H., Bull, J. M. & Ahmed, S. Optimal X-ray micro-CT image based methods for porosity and permeability quantification in heterogeneous sandstones. *Geophys J Int* **223**, 1210–1229 (2020).
22. Lucas-Oliveira, E. *et al.* Micro-computed tomography of sandstone rocks: Raw, filtered and segmented datasets. *Data in Brief* **41**, 107893 (2022).
23. Liu, J. (刘继龙), Xie, R. (谢然红), Guo, J. (郭江峰), Xu, C. (徐陈昱) & Wei, H. (卫弘媛). Multicomponent digital core construction and three-dimensional micro-pore structure characterization of shale. *Physics of Fluids* **35**, 082003 (2023).
24. Esteves Ferreira, M. *et al.* Full scale, microscopically resolved tomographies of sandstone and carbonate rocks augmented by experimental porosity and permeability values. *Sci Data* **10**, 368 (2023).
25. Ishola, O. & Vilcáez, J. Augmenting X-ray micro-CT data with MICP data for high resolution pore-scale simulations of flow properties of carbonate rocks. *Geoenergy Science and Engineering* **239**, 212982 (2024).
26. Wetzel, M., Kempka, T. & Kühn, M. Diagenetic Trends of Synthetic Reservoir Sandstone Properties Assessed by Digital Rock Physics. *Minerals* **11**, (2021).
27. Ibrahim, E. R., Jouini, M. S., Bouchaala, F. & Gomes, J. Simulation and Validation of Porosity and Permeability of Synthetic and Real Rock Models Using Three-Dimensional Printing and Digital Rock Physics. *ACS Omega* **6**, 31775–31781 (2021).
28. Song, R. *et al.* 3D Printing of natural sandstone at pore scale and comparative analysis on micro-structure and single/two-phase flow properties. *Energy* **261**, 125226 (2022).
29. Bryant, S. & Blunt, M. Prediction of relative permeability in simple porous media. *Phys. Rev. A* **46**, 2004–2011 (1992).
30. Chi, P., Sun, J., Zhang, R., Luo, X. & Yan, W. Reconstruction of large-scale

anisotropic 3D digital rocks from 2D shale images using generative adversarial network. *Marine and Petroleum Geology* **170**, 107065 (2024).
31. Yan, W., Golsanami, N., Xing, H., Li, S. & Chi, P. A Rapid Reconstruction Method of 3D Digital Rock with Strong Pore Connectivity. *Pure Appl. Geophys.* **181**, 1601–1616 (2024).
32. Čapek, P. On the Importance of Simulated Annealing Algorithms for Stochastic Reconstruction Constrained by Low-Order Microstructural Descriptors. *Transp Porous Med* **125**, 59–80 (2018).
33. Song, S. An improved simulated annealing algorithm for reconstructing 3D large-scale porous media. *Journal of Petroleum Science and Engineering* **182**, 106343 (2019).
34. Gerke, K. M., Karsanina, M. V. & Mallants, D. Universal Stochastic Multiscale Image Fusion: An Example Application for Shale Rock. *Sci Rep* **5**, 15880 (2015).
35. Karsanina, M. V. & Gerke, K. M. Hierarchical Optimization: Fast and Robust Multiscale Stochastic Reconstructions with Rescaled Correlation Functions. *Phys. Rev. Lett.* **121**, 265501 (2018).
36. Xu, Z., Teng, Q., He, X., Yang, X. & Li, Z. Multiple-point statistics method based on array structure for 3D reconstruction of Fontainebleau sandstone. *Journal of Petroleum Science and Engineering* **100**, 71–80 (2012).
37. Zhang, T., Du, Y., Huang, T. & Li, X. GPU-accelerated 3D reconstruction of porous media using multiple-point statistics. *Comput Geosci* **19**, 79–98 (2015).
38. Wu, Y. *et al.* Reconstruction of 3D porous media using multiple-point statistics based on a 3D training image. *Journal of Natural Gas Science and Engineering* **51**, 129–140 (2018).
39. Xie, Q., Xu, J., Yuan, Y. & Niu, C. Quantitative Analysis for the Reconstruction of Porous Media Using Multiple-Point Statistics. *Geofluids* **2020**, 8844968 (2020).
40. Okabe, H. & Blunt, M. J. Pore space reconstruction using multiple-point statistics. *Journal of Petroleum Science and Engineering* **46**, 121–137 (2005).
41. Huang, T., Lu, D.-T., Li, X. & Wang, L. GPU-based SNESIM implementation for multiple-point statistical simulation. *Computers & Geosciences* **54**, 75–87 (2013).
42. Zhang, T., Du, Y., Huang, T. & Li, X. GPU-accelerated 3D reconstruction of porous media using multiple-point statistics. *Comput Geosci* **19**, 79–98 (2015).
43. Huang, Y., Xiang, Z. & Qian, M. Deep-learning-based porous media microstructure quantitative characterization and reconstruction method. *Phys. Rev. E* **105**, 015308 (2022).
44. Li, J., Teng, Q., Zhang, N., Chen, H. & He, X. Deep learning method of stochastic reconstruction of three-dimensional digital cores from a two-dimensional image. *Phys. Rev. E* **107**, 055309 (2023).
45. Zhang, T., Shen, T., Hu, G., Lu, F. & Du, X. Stochastic reconstruction of digital cores using two-discriminator VAE-GAN. *Geoenergy Science and Engineering* **236**, 212744 (2024).
46. Li, Q. *et al.* Reconstructing the 3D digital core with a fully convolutional neural network. *Appl. Geophys.* **17**, 401–410 (2020).
47. Sepúlveda, E., Dowd, P. A. & Xu, C. Fuzzy Clustering with Spatial Correction

and Its Application to Geometallurgical Domaining. *Math Geosci* **50**, 895–928 (2018).
48. Wang, M., Wang, J., Pan, N. & Chen, S. Mesoscopic predictions of the effective thermal conductivity for microscale random porous media. *Phys. Rev. E* **75**, 036702 (2007).
49. Wang, M., Pan, N., Wang, J. & Chen, S. Mesoscopic simulations of phase distribution effects on the effective thermal conductivity of microgranular porous media. *Journal of Colloid and Interface Science* **311**, 562–570 (2007).
50. Huang, H., Chen, R., Tao, S., Wang, Y. & Zhang, L. Study of soot dynamic behavior and catalytic regeneration in diesel particulate filters. *Chemical Engineering Journal* **489**, 151498 (2024).
51. Xia, H. (夏胡熙), Lai, Y. (赖远明) & Mousavi-Nezhad, M. Meso-scale investigation on the permeability of frozen soils with the lattice Boltzmann method. *Physics of Fluids* **36**, 093616 (2024).
52. Li, T., Li, M., Jing, X., Xiao, W. & Cui, Q. Influence mechanism of pore-scale anisotropy and pore distribution heterogeneity on permeability of porous media. *Petroleum Exploration and Development* **46**, 594–604 (2019).
53. Zhao, D. *et al.* Numerical Study on Permeability of Reconstructed Porous Concrete Based on Lattice Boltzmann Method. *Buildings* **14**, (2024).
54. Ju, L., Wang, S., Shan, B., Sun, S. & Yan, B. A pore-scale lattice Boltzmann model for solute transport coupled with heterogeneous surface reactions and mineral dissolution. *Chemical Engineering Journal* **498**, 155191 (2024).
55. Hao, H. & Xu, Z. G. Pore-scale investigation on porous media morphology evolution considering dissolution and precipitation. *International Journal of Multiphase Flow* **168**, 104569 (2023).
56. Chen, S., Ai, B., Li, Y., Huang, X. & Yang, X. The effective thermal conductivity of random isotropic porous media analysis and prediction. *Engineering Analysis with Boundary Elements* **167**, 105895 (2024).
57. Yang, G., Liu, T., Lu, X. & Wang, M. Fast-QSGS: A GPU accelerated program for structure generation of granular disordered media. *Computer Physics Communications* **302**, 109241 (2024).
58. Harris, C. R. *et al.* Array programming with NumPy. *Nature* **585**, 357–362 (2020).
59. Okuta, R., Unno, Y., Nishino, D., Hido, S. & Loomis, C. CuPy: A NumPy-Compatible Library for NVIDIA GPU Calculations. *NIPS* (2017).
60. Valvatne, P. H. & Blunt, M. J. Predictive pore-scale modeling of two-phase flow in mixed wet media. *Water Resources Research* **40**, (2004).
61. Dong, H. & Blunt, M. J. Pore-network extraction from micro-computerized-tomography images. *Phys. Rev. E* **80**, 036307 (2009).
62. Bourbie, T. & Zinszner, B. Hydraulic and acoustic properties as a function of porosity in Fontainebleau Sandstone. *Journal of Geophysical Research: Solid Earth* **90**, 11524–11532 (1985).
63. Coker, D. A., Torquato, S. & Dunsmuir, J. H. Morphology and physical properties of Fontainebleau sandstone via a tomographic analysis. *Journal of Geophysical Research: Solid Earth* **101**, 17497–17506 (1996).

64. Saadi, F. A., Wolf, K.-H. & Kruijsdijk, C. V. Characterization of Fontainebleau Sandstone: Quartz Overgrowth and its Impact on Pore-Throat Framework. *J Pet Environ Biotechnol* **08**, (2017).
65. Doyen, P. M. Permeability, conductivity, and pore geometry of sandstone. *Journal of Geophysical Research: Solid Earth* **93**, 7729–7740 (1988).
66. Gomez, C. T., Dvorkin, J. & Vanorio, T. Laboratory measurements of porosity, permeability, resistivity, and velocity on Fontainebleau sandstones. *Geophysics* **75**, E191–E204 (2010).
67. Revil, A., Kessouri, P. & Torres-Verdín, C. Electrical conductivity, induced polarization, and permeability of the Fontainebleau sandstone. *Geophysics* **79**, D301–D318 (2014).